# Localized electronic states at grain boundaries on the surface of graphene and graphite

Adina Luican-Mayer[1], Jose E. Barrios-Vargas[2], Jesper Toft Falkenberg[3], Gabriel Autès[4], Aron W. Cummings[2], David Soriano[2], Guohong Li[1], Mads Brandbyge[3], Oleg V. Yazyev[4], Stephan Roche[2,5], and Eva Y. Andrei[1]

[1]Department of Physics and Astronomy, Rutgers University, Piscataway, New Jersey 08854, USA
[2]Catalan Institute of Nanoscience and Nanotechnology (ICN2), CSIC and The Barcelona Institute of Science and Technology, Campus UAB, Bellaterra, 08193 Barcelona, Spain
[3] Department of Micro- and Nanotechnology, Center for Nanostructured Graphene(CNG),Technical University of Denmark, Ørsteds Plads, Bldg. 345C, DK-2800, Kongens Lyngby, Denmark
[4]Institute of Physics, Ecole Polytechnique Fédérale de Lausanne (EPFL), CH-1015 Lausanne, Switzerland
[5]ICREA, Institució Catalana de Recerca i Estudis Avançats, 08070 Barcelona, Spain

Email: luican-mayer@uottawa.ca

Recent advances in large-scale synthesis of graphene and other 2D materials have underscored the importance of local defects such as dislocations and grain boundaries (GBs), and especially their tendency to alter the electronic properties of the material. Understanding how the polycrystalline morphology affects the electronic properties is crucial for the development of applications such as flexible electronics, energy harvesting devices or sensors. We here report on atomic scale characterization of several GBs and on the structural-dependence of the localized electronic states in their vicinity. Using low temperature scanning tunneling microscopy (STM) and spectroscopy (STS), together with tight binding and ab initio numerical simulations we explore GBs on the surface of graphite and elucidate the interconnection between the local density of states (LDOS) and their atomic structure. We show that the electronic fingerprints of these GBs consist of pronounced resonances which, depending on the relative orientation of the adjacent crystallites, appear either on the electron side of the spectrum or as an electron-hole symmetric doublet close to the charge neutrality point. These two types of spectral features will impact very differently the transport properties allowing, in the asymmetric case to introduce transport anisotropy which could be utilized to design novel growth and fabrication strategies to control device performance.

## 1. Introduction

Topological defects such as dislocations and GBs are inherently present, due to growth methods, in any crystalline material. Early STM experiments on graphite have identified the wire-like structures often observed on the surface, with the boundaries between misoriented grains of the crystal [1,2,3,4,5]. These studies focused on the topographical features associated with GBs, but it was only with the recent interest in graphene that theoretical [6,7,8] and experimental work [9,10,11][12,13,14,15] addressed questions regarding the electronic properties of GBs (for a review, see Ref. 16). Efforts to grow wafer-size graphene by chemical vapor deposition (CVD) revealed that the films are polycrystalline, consisting of micron size randomly oriented single crystal domains [17,18,19,20]. The GBs between adjacent domains are clearly observed in STM and in STS [9, 10,21,22,23], but the connection between the morphology of a GB and its effect on the associated local spectrum is still under debate [24]. In this work we combine structural and spectral characterization of GBs with *ab initio* and tight-binding calculations to show that GBs leave a strong imprint in the LDOS which depends on the misorientation angle between adjacent grains.

## 2. Results and discussion

We employed STM topography to identify GBs and subsequently STS to study their effect on the LDOS. Topographic surface maps show the GBs as narrow "wires" within which the lattice is reconstructed into a periodic pattern whose period is determined by the relative orientation between adjacent grains. The STM topography image showing the atomic structure of such a GB in figure 1(a), reveals a periodic pattern from which we can infer the misorientation angle between the grains, $\theta = 2\cdot\arcsin(b/2d)$, where $d$ is the period of the pattern and $b$ is the length of the Burgers vector of the dislocation forming the GB [25]. In the simplest possible situation, there is one dislocation per period with the length of the Burgers vector $b$ being equal to the lattice constant of graphene $a = 0.246$ nm, [6]. Thus, the period increases monotonically with decreasing misorientation angle. The angle was confirmed also by directly measuring the relative orientation of the crystallographic planes across the GB in the following way similarly to [6]: a vertical axis is drawn along the boundary and the corresponding horizontal axis is perpendicular to it. The sum of the angle corresponding to the grains to the left and right with respect to the horizontal axis are $\theta_L$ and $\theta_R$ respectively. The angle that will characterize the boundary is: $\theta = \theta_L + \theta_R$.

For the GB in figure 1(a) we find d = 0.6 nm, which corresponds closely to $\theta = 21.8°$ GB. The measured relative orientation of the two grains with respect to the boundary was, consistently, $\theta_L \sim \theta_R \sim 10.5°$. The atomic scale model of the $\theta = 21.8°$ GB is shown in figure 1(b). Figure 1(c) shows the simulated STM map calculated from first principles (see the Methods section below) for this GB. The dI/dV spectra far from this GB (figure 1(d)) display a V shape typical of the graphite surface, while on the GB a strong peak appears on the electron side at ~500 meV. Another common GB on the surface of graphite is shown in the atomically resolved STM image of figure 1(e). From the periodic structure of this GB, $d = 0.9$ nm, we infer that the observed GB is characterized by misorientation angle $\theta = 32.2°$. We find that the measured relative orientation of the two grains with respect to the boundary, $\theta_L \sim \theta_R \sim 16.5°$, is in good agreement. The atomic structure of this large-angle GB, schematically presented in figure 1(f), and its simulated STM image in figure 1(g) show the same qualitative features as the experimental data. We note that for this specific GB structure the relation between the misorientation angle $\theta$ and periodicity $d$ does not hold. This is due to the fact that this GB has two dislocations per period with differently oriented Burgers vectors. Hence, a modified expression $\theta = 60° - 2\cdot\arcsin(b_{tot}/2d)$ must be used [6], with $b_{tot}$ being the length of the vector resulting from summing the Burgers vectors of the two dislocations ($b_{tot} = \sqrt{3}a = 0.426$ nm in this case). The spectroscopic features associated with this GB, shown in figure 1(h), resemble those of the $\theta = 21.8°$ GB. In both cases the graphite surface to the left and right of this GB displays a V-shape spectrum whereas the spectrum measured on the GB has a distinctive peak at ~ 500 meV.

The observation of these two well-ordered GB structures is not accidental. These structural patterns correspond to the lowest formation energy for large-angle GBs in graphene [6] and are commonly observed in annealed Cu-grown epitaxial graphene [26] and in epitaxial graphene grown on C-face of SiC [23]. Remarkably, in spite of these structures being macroscopic in length their influence on the spectrum is localized to the immediate vicinity of the GB reflecting their true 1D nature. This is illustrated in figure 2 where we show the spatial distribution of the electronic states associated with the $\theta = 21.8°$ GB. The inherently short range nature of the GB's influence on the electronic structure is seen in the rapid decay (within ~2 nm) of the 500 meV peak, figures 2(a) and 2(b). Indeed, theoretical predictions suggest that the peak should decay exponentially away from the GB [8], in agreement with our observations. This is also consistent with the characteristic nanometer-scale decay lengths observed at the edge states in graphene [27,28].

We further investigate the electronic signature of the $\theta = 21.8°$ and $\theta = 32.2°$ GBs using density functional theory (DFT) and nearest-neighbors tight-biding (TB) electronic structure calculations in a monolayer lattice (see Methods). The calculated LDOS of the GBs are shown in figures 3(a) and 3(c) with black dashed and solid lines for DFT and TB, respectively. Both calculations exhibit a van Hove singularity (VHS) peak at positive energies which we compare to the peaks in the experimental $dI/dV$ curves (figure 1(d) and 1(h)). The position of the positive energy VHS predicted with TB for the $\theta = 21.8°$ GB closely reproduces the experimental feature. Similarly, a single VHS is predicted for the $\theta = 32.2°$ GB at positive energies, but its position does not show quantitative agreement with the experimental $dI/dV$ curve, indicating the need to refine the calculation as discussed below.

GBs are known to accumulate or redistribute local charges [29], for example as a result of interlayer interactions, which could change the electronic structure. We introduce the effect of charge redistribution in the TB Hamiltonian through an on-site energy $V_i = \omega \cdot \Delta\rho_i$ where $\omega$ is a weight factor and $\Delta\rho_i$ is excess charge (in units of electrons per carbon atom) on GB site $i$ obtained by

$$\Delta\rho_i = 1 - \int_{-\infty}^{E_F} LDOS_i(E) dE, \qquad (1)$$

where $LDOS_i(E)$ is the local density of states on site $i$ at energy $E$, and $E_F$ is the Fermi energy. The excess charge, Eq. (1), is shown in figure 3(b) and 3(d) for the $\theta = 21.8°$ and $\theta = 32.2°$ GBs respectively. The positive charge tends to be localized on the heptagons while the negative charge becomes more localized on the pentagons, which induces small dipoles along the GB. By introducing this approach, we were able to calculate the LDOS of both GBs (red line in figure 3(a) and figure 3(c)) including charge redistribution effects. It is worth noting that positive $\omega$ values shift the VHS on the electron side away from the charge neutrality point. In particular, setting the values $\omega = -5t$ and $\omega = -10t$ (with $t = -2.9$ eV being the hopping parameter) for the $\theta = 21.8°$ and $\theta = 32.2°$ GBs, respectively, we find that the calculated energies of the VHS closely matches the experimental values as shown in figures 3(a) and 3(c). For these values of $\omega$ the variation of the on-site energies is in the range [-0.26, 0.32] eV and [-2.0, 1.7] eV for the GBs with $\theta = 21.8°$ and $\theta = 32.2°$ respectively. These results highlight the importance of considering charge effects in graphene and graphite GBs. Additionally, for both GBs, DFT and TB calculations reveal a VHS at large negative energies that, being outside the experimental energy range, cannot be compared to the data. The energy value for the VHS in DFT calculations is showed in figure 3(a) and figure 3(c); meanwhile, in TB calculations the VHS is around 800 meV (See figure S3(c) in the supplementary information).

The STM topography images in figures 4(a), 4(b) correspond to a GB with an estimated θ ≈ 10° ± 6° and figure 4(c) shows its dI/dV spectrum. Such a small-angle GB corresponds to a large distance, d, between constituent dislocations and is not as easy to characterize numerically as the large-angle GBs discussed above. This is because, as detailed in the supplementary information (SI), in order to avoid interactions between periodic images of GBs in the numerical calculations one has to increase the sample size so that the distance between neighboring GBs is at least 4d. In figure 4(d) we show the LDOS for three small-angle GBs, θ = 3.5°, 4.4° and 7.2°, obtained by using TB without on-site corrections. The case with θ = 3.5° (d ~ 4 nm), where the spectrum exhibits two VHS peaks located on each side of the Dirac point and separated by 110 meV, gives the best fit with the experimental data. Interestingly, and in contrast to the θ = 21.8° and θ = 32.2° large-angle GBs, the two VHS peaks are more symmetric with respect to the Dirac point. Electron-hole symmetry in graphene is related to sublattice symmetry, and small angle GBs contain fewer defects per unit length, which translates into smaller degree of sublattice symmetry breaking. In figure 4(e) we compare the LDOS obtained with and without on-site charge corrections (red and black lines, respectively) for θ = 3.5° and ω = 10t. When the charge effects are considered both VHS peaks shift in the same direction so that their energies at ±50 meV correspond to perfect electron-hole symmetry in good agreement with the experimental results. As before, this indicates a clear impact of charge distribution on the exact position of the VHS peaks.

In addition to the energetically favorable cases described above, we find that in fact the most abundant GBs on the surface of graphite are irregular as the one shown in figures 4(g) and 4(h), which appear as a zigzag pattern. Such GBs, referred to as amorphous, are characterized by the presence of additional disorder (such as irregular distances between the constituent dislocations and relaxation of the lattice in the GB) beyond the simple model of a periodic array of dislocations [30]. The relaxation of the lattice allows for an extension of the available orientations of the grains. Furthermore, we observed that imaging the atomic details of such a GB along its entire length is challenging. While some segments of the GB are free of defects and produce clear topographic maps, others appear blurry indicating that the tunneling junction is not stable when the tip is scanned across them. Such instability could be the result of dangling bonds, as GBs have an enhanced chemical reactivity [31]. The dI/dV maps in figure 4(i) clearly show the existence of localized electronic states at the GB, but their energies vary from segment to segment as a consequence of the irregular structure of the GB.

## 3. Conclusion

Our experimental and numerical results show that it is possible to correlate the spectroscopic signature of a GB with its atomic structure. The GBs induce 1D localized electronic states with fingerprint spectroscopic resonances reflecting the misalignment angle between adjacent grains. For GBs characterized by small misorientation angles the spectra exhibit a pair of electron-hole symmetric VHS peaks close to the charge neutrality point, while at larger misorientation angles the electron-hole symmetry is broken. The strong sensitivity of the spectral response to the morphology of GBs and, in particular the electron-hole spectral asymmetry emerging at large angles, suggest a pathway to tailor and control the electronic response of graphene by GB engineering. Furthermore, these results can aid in developing a deeper understanding of the physical and electronic structure of polycrystalline graphene, and ultimately to optimize growth and fabrication protocols to control the effects of GBs on its electrical transport properties.

## 4. Methods

**STM and STS measurements.**

The STM tips used in this experiment were mechanically cut from Pt-Ir wire. The tunneling conductance dI/dV was measured using lock-in detection at 340 Hz. The samples were obtained from highly oriented pyrolitic graphite cleaved in air and immediately transferred to the STM. Many GBs imaged on the surface of graphite are not restricted to the top layer. GBs in the top layer are accompanied by a Moiré pattern since the stacking in this case is no longer Bernal between the top and second layer [31]. The absence of such a Moiré pattern indicates the presence of a plane of GBs penetrating into the graphite crystal.

**Electronic structure calculations.**

First-principles calculations were performed within the DFT framework employing the generalized gradient approximation (GGA) and the Perdew-Burke-Ernzerhof (PBE) exchange-correlation functional. We used the double-zeta plus polarization (DZP) localized-orbital basis set employing the SIESTA package [32]. The GB models were described by two-dimensional periodic supercells containing two complimentary GBs. The $\theta = 21.8°$ GB supercell model of dimensions 0.65 nm × 4.23 nm contained 100 atoms. The $\theta = 32.2°$ GB supercell model of dimensions 0.89 nm × 4.18 nm contained 140 atoms. These structures were relaxed down to 0.04 eV/Å maximum forces with variable cell using a $12 \times 2$ $k$-point mesh. The electronic structure calculations were then performed self consistently with $144 \times 6$ and $72 \times 6$ $k$-point meshes, respectively. The DFT results were compared to a one-orbital tight-binding (TB) model

$$\hat{H} = -t\sum_{\langle ij \rangle}\left(c_i^\dagger c_j + h.c.\right) + \sum_{i \in GB} V_i c_i^\dagger c_i,$$

where $c_i^\dagger$ and $c_j$ are creation and annihilation operators, respectively, and $t = 2.9$ eV is nearest-neighbor hopping parameter. The second term accounts for the charge variations taking place at the grain boundary via the on-site energies $V_i$. The TB LDOS curves were computed using a Lanczos recursion method. Both DFT and TB calculations were performed using wide enough supercells in order to avoid inter-grain interactions (See Supplementary Information for details).

References:

[1] C. R. Clemmer and T. P. Beebe Jr. Graphite: a mimic for dna and other biomolecules in scanning tunneling microscope studies. Science, 251(4994):640–642, 1991.

[2] W. M. Heckl and G. Binnig. Domain walls on graphite mimic DNA. Ultramicroscopy, 42:1073–1078, 1992.

[3] C. Daulan, A. Derré, S. Flandrois, J. Roux, H. Saadaoui, et al. STM observations at the atomic scale of a tilt grain sub-boundary on highly oriented pyrolytic graphite. Journal de Physique I, 5(9):1111–1117, 1995.

[4] T. R. Albrecht, H. A. Mizes, J. Nogami, S. Park, and C. F. Quate. Observation of tilt boundaries in graphite by scanning tunneling microscopy and associated multiple tip effects. Applied Physics Letters, 52(5):362–364, 1988.

[5] P. Simonis, C. Goffaux, P. A. Thiry, L. P. Biro, P. Lambin, and V. Meunier. STM study of a grain boundary in graphite. Surface Science, 511(1):319–322, 2002.

[6] O. V. Yazyev and S. G. Louie. Topological defects in graphene: Dislocations and grain boundaries. Physical Review B, 81(19):195420, 2010.

[7] Y. Liu, B. I. Yakobson. Cones, Pringles, and Grain Boundary Landscapes in Graphene Topology. Nano Lett., 10:2178–2183, 2010.

[8] A. Mesaros, S. Papanikolaou, C. F. J. Flipse, D. Sadri, and J. Zaanen. Electronic states of graphene grain boundaries. Physical Review B, 82(20):205119, 2010.

[9] J. Červenka and C. F. J. Flipse. Structural and electronic properties of grain boundaries in graphite: Planes of periodically distributed point defects. Phys. Rev. B, 79:195429, 2009.

[10] J. Červenka, M. I. Katsnelson, and C. F. J. Flipse. Room-temperature ferromagnetism in graphite driven by two-dimensional networks of point defects. Nature Physics, 5(11):840–844, 2009.

[11] J. Lahiri, Y. Lin, P. Bozkurt, I. I. Oleynik, and M. Batzill. An extended defect in graphene as a metallic wire. Nature Nanotechnology, 5(5):326–329, 2010.

[12] O. V. Yazyev and S. G. Louie. Electronic transport in polycrystalline graphene. Nature Materials, 9(10):806–809, 2010.

[13] D. Gunlycke and C. T. White. Graphene valley filter using a line defect. Physical Review Letters, 106(13):136806, 2011.

[14] Q. Yu, L. A. Jauregui, W. Wu, R. Colby, J. Tian, Z. Su, H. Cao, Z. Liu, D. Pandey, D. Wei, et al. Control and characterization of individual grains and grain boundaries in graphene grown by chemical vapour deposition. Nature Materials, 10(6):443–449, 2011.

[15] F. Gargiulo and O. V. Yazyev. Topological Aspects of Charge-Carrier Transmission across Grain Boundaries in Graphene. Nano Lett. 14:250–254, 2014.

[16] O. V. Yazyev and Y. P. Chen, Polycrystalline graphene and other two-dimensional materials, Nature Nanotechnology, 9:755–767, 2014.

[17] P. Y. Huang, C. S. Ruiz-Vargas, A. M. van der Zande, W. S. Whitney, M. P. Levendorf, J. W. Kevek, S. Garg, J. S. Alden, C. J. Hustedt, Y. Zhu, et al. Grains and grain boundaries in single-layer graphene atomic patchwork quilts. Nature, 469(7330):389–392, 2011.

[18] K. Kim, Z. Lee, W. Regan, C. Kisielowski, M. F. Crommie, A. Zettl, Grain Boundary Mapping in Polycrystalline Graphene, ACS Nano, 5:2142–2146, 2011.

[19] J. An, E. Voelkl, J. W. Suk, X. Li, C. W. Magnuson, L. Fu, P. Tiemeijer, M. Bischo, B. Freitag, E. Popova, et al. Domain (grain) boundaries and evidence of twinlike structures in chemically vapor deposited grown graphene. ACS Nano, 5(4):2433, 2011.

[20] J. H. Chen, G. Autès, N. Alem, F. Gargiulo, A. Gautam, M. Linck, C. Kisielowski, O. V. Yazyev, S. G. Louie, A. Zettl, Controlled growth of a line defect in graphene and implications for gate-tunable valley filtering, Phys. Rev. B, 89:121407, 2014.

[21] J. C. Koepke, J. D. Wood, D. Estrada, Z.-Y. Ong, K. T. He, E. Pop, and J. W. Lyding. Atomic-Scale Evidence for Potential Barriers and Strong Carrier Scattering at Graphene Grain Boundaries: A Scanning Tunneling Microscopy Study. ACS Nano 7(1):75–86, 2013.

[22] C. Ma, H. Sun, Y. Zhao, B. Li, Q. Li, A. Zhao, X. Wang, Y. Luo, J. Yang, B. Wang, J.G. Hou. Evidence of van Hove singularities in ordered grain boundaries of graphene. Phys. Rev. Lett. 112:226802, 2014.

[23] Y. Tison, J. Lagoute, V. Repain, C. Chacon, Y. Girard, F. Joucken, R. Sporken, F. Gargiulo, O. V. Yazyev, S. Rousset. Grain Boundaries in Graphene on SiC(000$\bar{1}$) Substrate. Nano Lett., 14:6382–6386, 2014.

[24] A. W. Cummings, D. L. Duong, V. L. Nguyen, D. V. Tuan, J. Kotakoski, J. E. Barrios-Vargas, Y. H. Lee, S. Roche. Charge transport in polycrystalline graphene: challenges and opportunities. Adv. Mater. 26:5079–5094, 2014.

[25] J. P. Hirth, J. Lothe. Theory of Dislocations. Wiley, New York, 1982.

[26] B. Yang, H. Xu, J. Lu, K. P. Loh. Periodic Grain Boundaries Formed by Thermal Reconstruction of Polycrystalline Graphene Film. J. Am. Chem. Soc., 136:12041-12046, 2014.

[27] C. Tao, L. Jiao, O.V. Yazyev, Y.-C. Chen, J. Feng, X. Zhang, R.B. Capaz, J.M. Tour, A. Zettl, S.G. Louie, H. Dai, M.F. Crommie, Spatially resolving edge states of chiral graphene nanoribbons, Nature Physics, 7:616-620, 2011.

[28] G. Li, A. Luican-Mayer, D. Abanin, L. Levitov, E.Y. Andrei, Evolution of Landau levels into edge states in graphene, Nature Communications, 4:1744, 2013.

[29] S. Ihnatsenka and I. V. Zozoulenko. Electron interaction, charging, and screening at grain boundaries in graphene. Phys. Rev. B, 88:085436, 2013.

[30] W. T. Read, W. Shockley. Dislocation Models of Crystal Grain Boundaries. Phys. Rev., 78:275, 1950.

[31] B. Wang, Y. Puzyrev, and S. T. Pantelides. Strain enhanced defect reactivity at grain boundaries in polycrystalline graphene. Carbon, 49(12):3983–3988, 2011.

[32] J. M. Soler, E. Artacho, J. D. Gale, A. García, J. Junquera, P. Ordejón, D. Sánchez-Portal. The SIESTA method for ab-initio order-N materials simulation. J. Phys.: Condens. Matt. 14:2745–2779, 2002.

[31] Li, G. Li, A. Luican, J. M. B. L. dos Santos, A. H. Castro Neto, A. Reina, J. Kong, E.Y. Andrei. Observation of Van Hove singularities in twisted graphene layers. Nature Physics, 6:109-113, 2010.

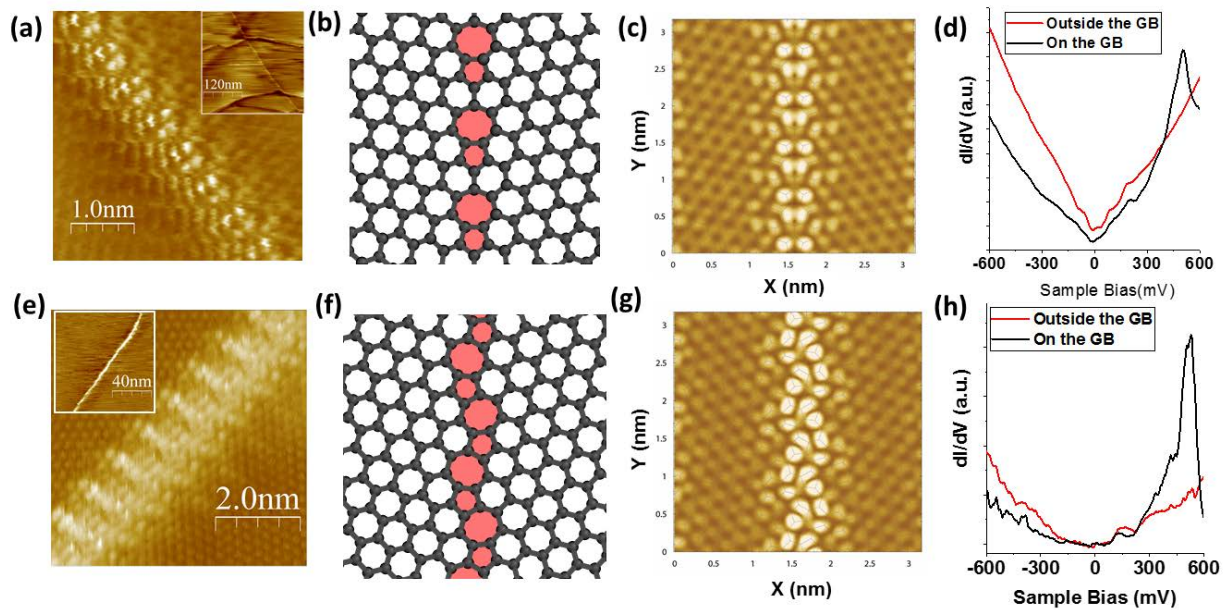

**Figure 1. STM and STS characterization of GBs on the graphite surface.** (a) STM topography of the θ = 21.8° GB. The main panel shows the atomically resolved image of the periodic structure, d=0.6nm, associated with this GB. The inset shows a large scale view of the same GB. (Scanning parameters: $V_{bias}$=600mV; $I_{tunnel}$=22pA). (b) Atomic-scale model showing the 5-7 lattice reconstruction at the θ = 21.8° GB site. The 5- and 7-membered rings are highlighted. (c) Simulated STM maps calculated from first principles by integrating the LDOS from 0 to 600meV. (d) Comparison of STS measured on the θ = 21.8° GB (black) and on the neighboring graphite surface (red). (e) Same as (a) for the θ = 32.2° (d = 0.9 nm) GB. ($V_{bias}$=500 mV; $I_{tunnel}$=21pA ). (f) Atomic-scale model showing the 5-7 lattice reconstruction at the θ = 32.2° GB site. The 5- and 7-membered rings are highlighted. (g) Simulated STM maps calculated from first principles by integrating the LDOS from 0 to 500meV. (h) Same as (d) for the θ = 32.2° GB.

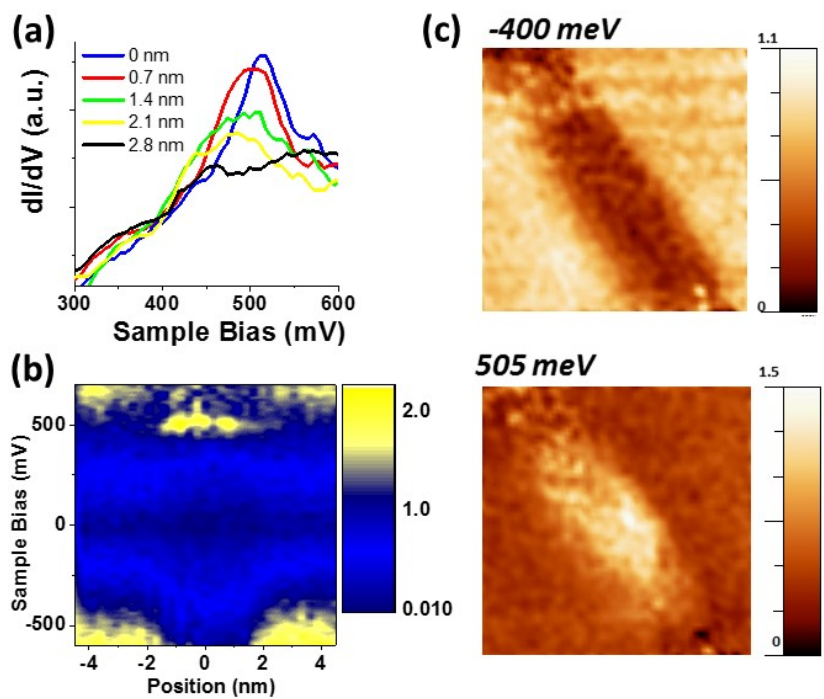

**Figure 2. Localized electronic states at a GB.** (a) STS at the different positions, perpendicular to the θ = 21.8° GB (0nm represents the center of the GB). (b) dI/dV map along a perpendicular line to the GB in figure 1(a). (c) dI/dV maps (10nm ×10nm) at the energies indicated in each panel. The bright color represents a high value for dI/dV.

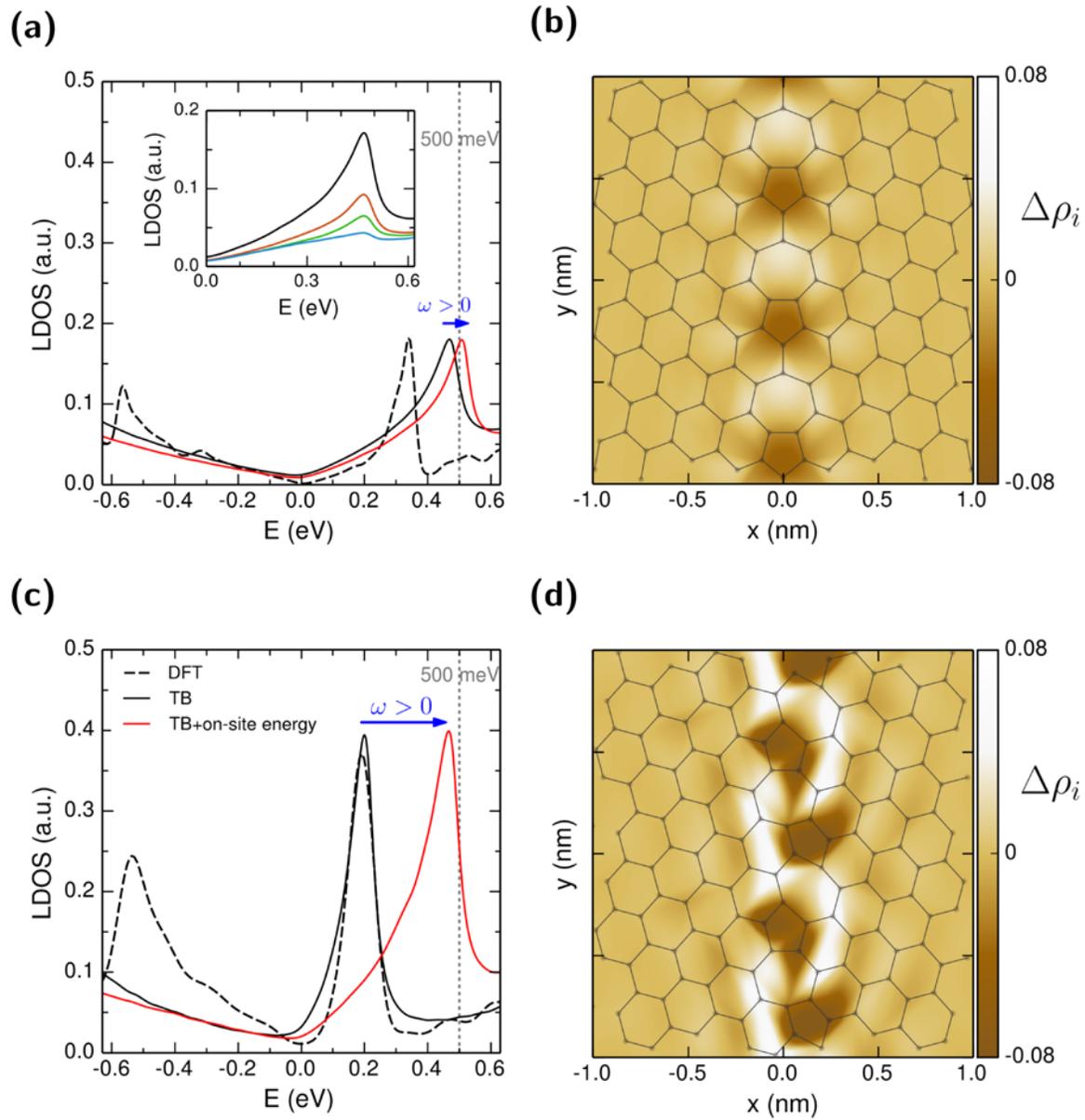

**Figure 3: TB and DFT characterization of GBs.** LDOS at the GB of the (a) $\theta = 21.8°$ GB and (c) $\theta = 32.2°$. Black and red solid lines corresponds to TB calculations without and with on-site energy dependence, respectively. Dashed line corresponds to the LDOS from DFT calculations. The inset in panel (a) shows the decay of the peak intensity while moving away from the center of the GB in steps of 2.46 Å. The gray dashed line at 500 meV indicates the energy of the experimentally measured VHS. The arrow indicates the direction of the shift of the VHS in the presence of charge redistribution. Panels (b) and (d) show the charge distribution along the GB for the $\theta = 21.8°$ GB and $\theta = 32.2°$ GB, respectively.

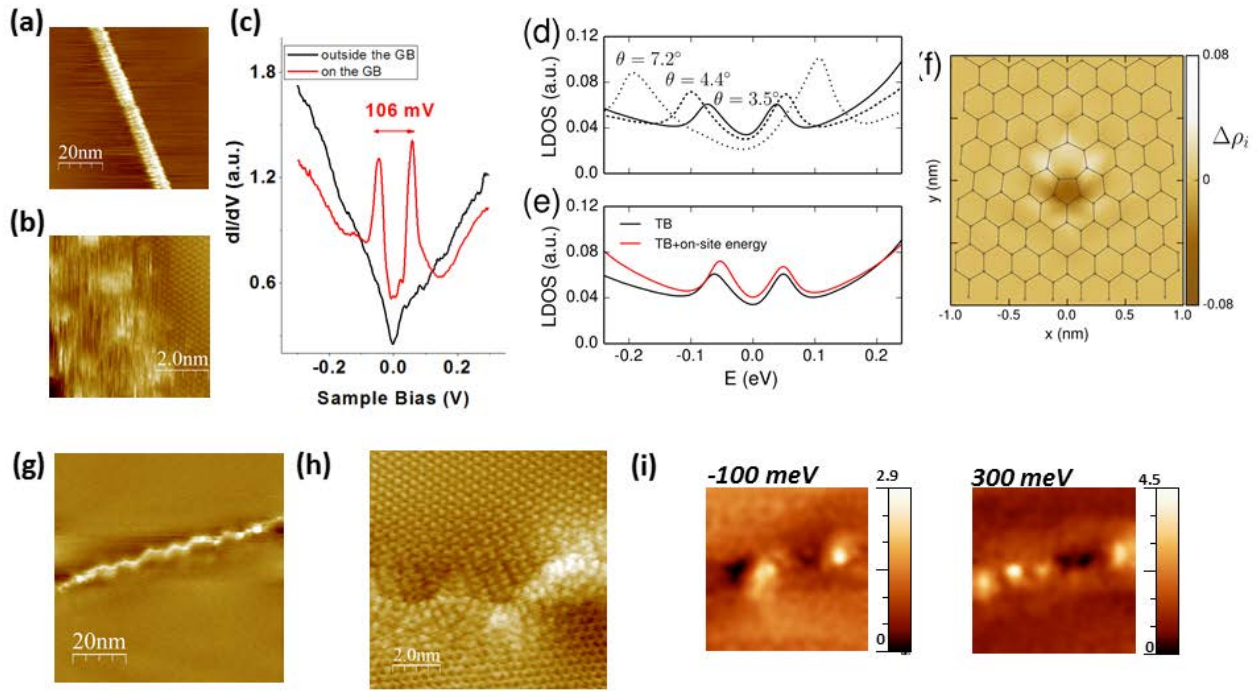

**Figure 4: Small angle and amorphous GBs** (a) Large area STM topographic images of the θ = (10 ±6) ° GB. (b) Atomic resolution STM of the GB in (a) showing a period of 2.5nm. (Scanning parameters: $V_{bias}$=300mV; $I_{tunnel}$=20pA). (c) dI/dV spectrum corresponding to this GB in (a). (d) Calculated LDOS at low angle GBs for θ =3.5°, 4.4° and 7.2° as indicated, obtained by using TB without on-site corrections. (e) TB calculated LDOS of the θ =3.5° GB without (black line) and with (red line) including the on-site charge distribution. (f) On site charge distribution along the θ =3.5° GB. (g) STM topographic image of an amorphous GB. (h) Atomically resolved segment of the GB in (g). (Scanning parameters: $V_{bias}$=300mV; $I_{tunnel}$=19pA). (i) dI/dV maps (30nm ×30nm) at the indicated energies showing how the different segments of the GB host electronic states at different energies.

# Supplementary Material

**CONTENTS**

**THE ANGLE OF A GRAIN BOUNDARY**

To characterize a GB between two graphene grains we use the following convention depicted in Figure S1: a vertical axis is drawn along the boundary and the corresponding horizontal axis is perpendicular to it. The angle corresponding to the grains to the left and right with respect to the horizontal axis $\theta_L$ and $\theta_R$ respectively. The angle that will characterize the boundary is: $\theta = \theta_L + \theta_R$.

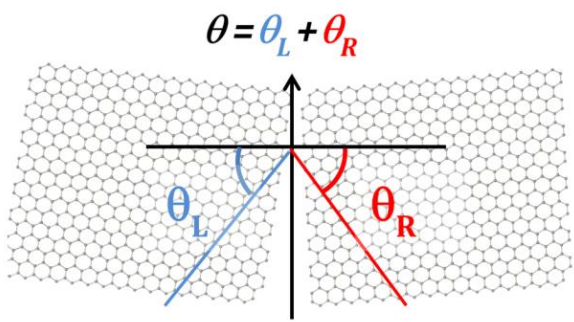

FIG. S1. Schematic of the convention for measuring the angle of a grain boundary.

**INCLUDING PERIODIC BOUNDARY CONDITIONS IN THE LATTICE**

A unit cell lattice with periodic boundary conditions is needed for DFT and tight binding calculations, meanwhile a tight binding calculation can be performed considering a grain boundary (GB) attached to two semiinfinite leads using the Landauer approach, as is described in Ref [1]. The calculated density of states curves for $\theta = 21.8°$ GB and $\theta = 32.2°$ are shown in Figs. S2A and B, respectively. The DFT and TB models provided similar results except for the more complicated structure of the density of states at negative energies in the case of $\theta = 32.2°$ GB. This discrepancy is attributed to the inter-GB interaction in the periodic model used in DFT calculations as we will show next.

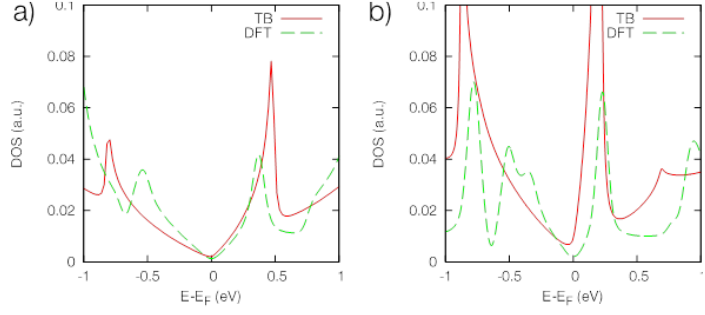

FIG. S2. Density of states of the (a) $\theta = 21.8°$ GB and (b) $\theta = 32.2°$ models obtained from DFT and tight binding calculations.

We implement this periodicity using an inverted copy of the GB separated a distance $\Delta x$ as is schematically shown in Figs. S3A and B. To confirm that there is not interference between the GB, we calculate the local density of states (LDOS) of GB as a function of $\Delta x$ using the tight binding Hamiltonian showed in the main text. It is helpful to define the ratio between the periodicity ($\Delta y \equiv d$) of the GB and $\Delta x$, $r_{GB} \equiv \Delta y/\Delta x$. As we can see in Figs. S3C and D, if $r_{GB} < 1/4$ the Van Hove singularities are no longer affected as much for tight binding calculations as for DFT ones.

## MOIRÉ PATTERNS DUE TO THE OVERLAP OF NEIGHBORING GRAPHENE GRAINS

Here we consider the possibility that the GB depicted in Figs. 4(a) and (b) of the main text is an overlap GB, where the two grains, instead of stitching together in-plane, lie on top of one another to form a narrow bilayer region. This type of overlapping structure has been observed in CVD-grown single-layer and multi-layer graphene [2, 3], and may also exist in graphite. The structure we consider is shown in Fig. S4A, where the red open (blue closed) symbols represent the carbon atoms of the top (bottom) layer of the overlap GB. The layers have been rotated 10.9° with respect to one another, and we consider an overlap region that is 10 nm wide and infinitely long. As can be seen in Fig. S4A, a rotation of 10.9° results in a Moiré pattern with a periodicity of ~1.5 nm. The top layer, being unrotated, is terminated with a zigzag edge, while the edge of the bottom layer is neither zigzag nor armchair due to its rotation angle. We describe the electronic structure of this system with a tight-binding approximation, using a parametrization that has been developed for twisted bilayer graphene [4].

In Fig. S4B we plot the total DOS of the top layer in the overlap region. For comparison, we also plot the DOS of 2D bilayer graphene. The 2D bilayer system shows the expected van Hove singularities arising from the periodicity of the Moiré pattern [4]. These features are also present in the overlap GB, but an additional set of peaks appears at low energies near the Dirac point, which qualitatively resemble those seen in the STS measurements. To investigate the origin of these peaks, in Fig. S4C we plot the LDOS of the top layer at three points - two in the middle of the overlap GB, marked 1 and 2 in Fig. S4A, and one at the zigzag edge of the top layer. At the zigzag edge, the LDOS is dominated by a pair of peaks near the Dirac point. These peaks correspond to localized edge states that are split by the interaction with the bottom layer. The splitting of these peaks is ~90 meV, close to what was measured with STS in Fig. 4(c) of the main text. However, these peaks do not appear at positions 1 and 2 in the middle of the overlap GB, contrary to the experiments, which reveal strong peaks everywhere within the GB. Finally, in Fig. S4D we map the LDOS of the top layer of the overlap GB at an energy of -0.21 eV, corresponding to one of the localized peaks at the zigzag edge. This LDOS map confirms a strong localization of this state at the zigzag edge. In addition, fluctuations are also present in the LDOS, with a periodicity matching that of the underlying Moiré pattern, and of the STM scan shown in the Fig. 3B.

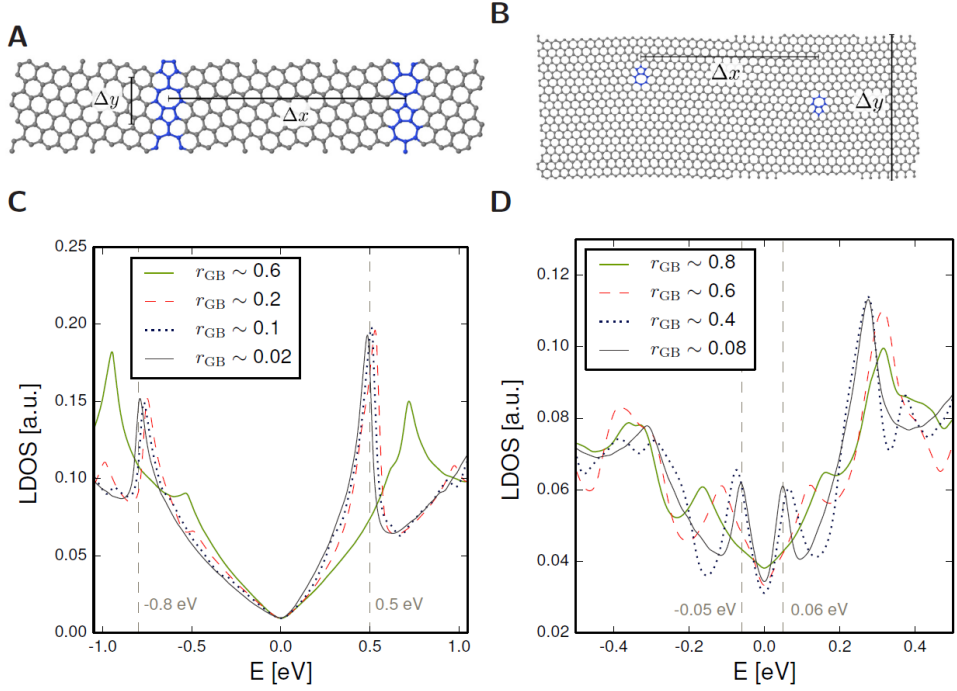

FIG. S3. (A) and (B) Ball and sticks sketches of the units cells typically used to evaluated LDOS, in which the vertices of the odd member rings are in blue. (C) and (D) Examples of LDOS for different ratios, $r_{GB} \equiv \Delta y/\Delta x$ keeping the GB periodicity fix ($\Delta y \equiv d$) where we can noticed that if $r_{GB} < 1/4$ the Van Hove singularities are well defined. In panels (A) and (C) the periodicity of the GB $\Delta y$ = 6.5 A meanwhile in panels (B) and (D) $\Delta y$ = 4nm.

The overlap GB provides several qualitative similarities to the experimental measurements of the GB in Fig. 4 of the main text, including the periodicity of the LDOS fluctuations and the appearance of two peaks near the Dirac point that are split by ~90 meV. However, these peaks only appear at the edge of the overlap GB, and rely on a very specific condition - a zigzag termination of the top graphene layer that can be probed by the STM. As with the structure proposed in Fig. 4(f) if the main text, it is therefore difficult to conclusively say that the 10° ± 6° is due to an overlap of neighboring graphene grains without conducting more detailed characterization of its atomic structure.

[1] F. Gargiulo and O. V. Yazyev, Nano Letters 14, 250 (2014).

[2] A. W. Tsen, L. Brown, M. P. Levendorf, F. Ghahari, P. Y. Huang, R. W. Havener, C. S. Ruiz-Vargas, D. A. Muller, P. Kim, and J. Park, Science 336, 1143 (2012).

[3] A. W. Robertson, A. Bachmatiuk, Y. A. Wu, F. Schffel, B. Rellinghaus, B. Bchner, M. H. Rmmeli, and J. H. Warner, ACS Nano 5, 6610 (2011).

[4] I. Brihuega, P. Mallet, H. González-Herrero, G. Trambly de Laissardìere, M. M. Ugeda, L. Magaud, J. M. Gomez-Rodrıguez, F. Yndurain, and J.-Y. Veuillen, Phys. Rev. Lett. 109, 196802 (2012).